\documentclass[preprint,showpacs,preprintnumbers,amsmath,amssymb]{revtex4}

\usepackage[dvips]{epsfig}

\begin{document}


\title{Using dileptons to probe the Color Glass Condensate}

\author{ M. A. Betemps$^{1,2}$ }
\email{marcos.betemps@ufrgs.br}
\author{M. B. Gay Ducati$^{1}$ }
\email{beatriz.gay@ufrgs.br}

\affiliation{
$^1$ High Energy Physics Phenomenology Group, GFPAE\\
Instituto de F\'{\i}sica, Universidade Federal do Rio Grande do Sul\\
Caixa Postal 15051, CEP 91501-970, Porto Alegre, RS, Brazil.\\
$^2$ Conjunto Agrot\'ecnio ``Visconde da Gra\c ca'', CAVG\\
Universidade Federal de Pelotas \\
Caixa Postal 460, CEP 96060-290, Pelotas, RS, Brazil.
}

\date{\today}

\begin{abstract}

The rapidity and transverse momentum dependence of the nuclear
modification ratio for dilepton production at RHIC and LHC is
presented, calculated in the Color Glass Condensate (CGC)
framework. The transverse momentum ratio is compared for two distinct
dilepton mass values and a suppression of the Cronin peak is verified
even for large mass. The nuclear modification ratio suppression in the
dilepton rapidity spectra, as obtained experimentally for hadrons at
RHIC, is verified for LHC energies at large transverse momentum,
although not present at RHIC energies. The ratio between LHC and RHIC
nuclear modification ratios is evaluated in the CGC, showing the large
saturation effects at LHC compared with the RHIC results. These
results consolidate the dilepton as a most suitable observable to
investigate the QCD high density approaches.

\end{abstract}


\pacs{ 11.15.Kc,  24.85.+p}


\maketitle

\section{Introduction} 

In the most recent data on hadrons transverse momentum spectra in  high energy collisions, some
interesting results have been pointed out
\cite{Debbe:2004ci,Arsene:2004ux}.  The comparison between
transverse momentum distribution in $d-Au$ and $p-p$, and in
central and peripheral collisions has being the main object of
investigation. These comparisons are performed introducing the nuclear
modification ratio, 
\begin{eqnarray}
R_{dAu}=\frac{\frac{dN^{dAu\rightarrow
hX}}{dydp_T^2}}{N_{coll}\frac{dN^{pp\rightarrow hX}}{dydp_T^2}}, 
\end{eqnarray}
where the normalization factor $N_{coll}$ is the number of binary
collisions, $y$ and $p_T$ are the rapidity and transverse momentum of
the hadrons, respectively.  For the case of $d-Au$ collisions at
mid-rapidity $y \approx 0$, the ratio $R_{dAu}$ becomes smaller than 1
at small $p_T$, larger than 1 at intermediated values of $p_T$ and
saturates at 1 for large $p_T$. This implies a peak at
intermediate $p_T$ (2-5 GeV), which is called Cronin effect or Cronin
peak.  At large rapidities a suppression of the ratio $R_{dAu}$ was
observed and a flat behavior at large $p_T$ is found (disappearance of
the Cronin peak). Another ratio should be evaluated, considering the
same process, now taking a ratio between central and peripheral
process in order to avoid systematic errors. In this case the ratio
$R_{CP}$ should be taken instead of the ratio $R_{dAu}$.  In both
defined ratios, the observations are the same, a significant reduction
of the yield of charged hadrons measured in $dAu$ collisions when
compared with the $pp$ collisions at forward rapidities
\cite{Arsene:2004ux}.

The main interest related to the Cronin effect in the literature is
regarding the hadron $p_T$ spectra, however, the dilepton production
should also be considered as an observable to study this effect. The
dileptons analyzed in the context of the Color Glass Condensate (
e.g. \cite{MABMBGDCGC1,Jalilian-Marian:2004er,Baier:2004tj,Jalilian-Marian:2005jf})
present the same features of the Cronin effect in such approach
\cite{Blaizot:2004wu} and do not present final state 
interactions, providing more clear information on the Color Glass
Condensate. In this work, the ratio between $p-Au$ and $p-p$
differential cross section is evaluated, analyzing the transverse
momentum distribution at a fixed rapidity and the rapidity
distribution of the dileptons for 6 GeV mass. A comparison
between distinct dilepton mass results in the transverse momentum
distribution at a fixed rapidity is performed.

The origin of the Cronin effect can be explained in different ways,
depending on the rapidity region. In $d-Au$ collisions and central
rapidities, the Cronin peak is due to the multiple scattering of the
deuterium constituents with the nuclear environment . At forward
rapidities the suppression of the ratio could be a manifestation of the CGC
\cite{CGCsatCron,Albacete:2003iq}  (or effect of $x_F\rightarrow 1$
\cite{Kopeliovich2}). At backward rapidities in $d-Au$ collisions, 
the nuclear modification ratio should present different behavior once
comparing with the hadron and dilepton $p_T$ spectra
\cite{Kopeliovich3}. Concerning dilepton at backward rapidities, this
point is being under investigation in another work \cite{MABMBGDnew}.

\section{Dilepton production in the CGC approach} 

The dilepton production
at high energies is dominated by the bremsstrahlung of a virtual
photon by a quark from a hadron interacting with a dense background
gluonic field of the nucleus, and afterwards decaying into a lepton
pair. In order to do the required calculation one considers only the diagrams where the
photon emission occurs before or after the interaction with the
nucleus, since the emission considering both before and after the
interaction is suppressed
\cite{Gelis:2002ki}.

The dilepton production is investigated in the context
of the Color Glass Condensate, which is a QCD classical effective
theory to deal with the high dense partonic system \cite{CGC}.  In
this theory, the small $x$ gluons are radiated from fast moving color
sources, which are partons with larger values of $x$, being
described by a color source density $\rho_a$ with internal dynamics
frozen by Lorentz time dilatation, thus forming a color glass. The
observables are obtained by means of an average over all
configurations of the color sources, performed through a weight
functional $W_{\Lambda^+}[\rho]$, which depends upon the dynamics of
the fast modes, and upon the intermediate scale $\Lambda^+$, which
defines fast ($p^+>\Lambda^+$) and soft ($p^+<\Lambda^+$) modes.  The
effective theory is valid for soft momenta of order $\Lambda^+$.
Indeed, reaching much softer scale, there are large radiative
corrections which invalidate the classical approximation. The
modifications to the effective classical theory are governed by a
functional, nonlinear, evolution equation JIMWLK
\cite{RGE1,RGE2} for the statistical weight functional
$W_{\Lambda^+}[\rho]$ associated with the random variable $\rho_a(x)$.

The hadronic cross section of the process is obtained employing the
collinear factorization and considering the forward rapidity region
\cite{MABMBGDCGC1,dileptonGelisJJI},
\begin{eqnarray}\nonumber
\frac{d\sigma^{pA\rightarrow
    ql^+l^-X}}{dp_T^2\,dM\,dy}&\!\!=&\!\!\frac{4\pi^2}{M} R^2_A
\frac{\alpha_{em}^2}{3\pi}\\
\times \int \frac{dl_T}{(2\pi)^3}&\!\!\!\!\! & \!\!\!\!\! l_T
\,    {\mathcal W}(p_T,l_T,x_1)\,C(l_T,x_2,A),
\label{eqcsh}
\end{eqnarray}
with $y$ being the rapidity, $s$ represents the squared center of mass
energy and $l_T$ is the total transverse momentum transfer
between the nucleus and the quark. $R_A$ is the nuclear radius, $M$ is
the lepton pair mass, $x_1$ and $x_2$ are the momentum fractions
carried by the quark from the proton and by the gluonic field from the
nucleus, respectively, defined in the formal way.  The expression
(\ref{eqcsh}) is restricted to the forward region only, which means
positive rapidities $y$.  The function ${\mathcal W}(p_T,l_T,x_1)$ can be
written as \cite{dileptonGelisJJI},
\begin{eqnarray}\nonumber
&&{\mathcal W}(p_T,l_T,x_1)=\int_{x_1}^{1}dz\,z  F_2 (x_1/z,M^2)\\\nonumber
&&\times \left\{
 \frac{(1+(1-z)^2)z^2 l_T^2}{[ p_T^2+M^2(1-z)][( p_T-z
    l_T)^2+M^2(1-z)]}\right. \\\nonumber
&\!\!-&\!\!z(1-z)M^2\left[\frac{1}{[ p_T^2+M^2(1-z)]}\right.\\
&\!\!-&\!\! \left.\left.\frac{1}{[(
    p_T-z l_T)^2+M^2(1-z)]}\right]^2 \right\},
\label{cs2}
\end{eqnarray}
which selects the values of $l_T$ larger than $p_T$
\cite{MABMBGDCGC1}. Here, $F_2(x_1/z,Q^2)$ is the partonic structure function,
which takes into account the quark distribution of the proton
projectile and $z\equiv p^-/k^-$ (light-cone variables) is the energy
fraction of the proton carried by the virtual photon.  In the
Eq. \ref{eqcsh} the function $C(l_T,x_2,A)$ is the field correlator
function which, disregarding the energy and nuclear dependence, can be
defined by
\cite{Gelisqq1},
\begin{eqnarray}
C(l_T)\equiv \int d^2 x_{\perp} e^{il_T\cdot
  x_{\perp}}\langle U(0)U^{\dagger}(x_{\perp})\rangle _{\rho},
\label{defCk}
\end{eqnarray}
with the averaged factor representing the average over all
configurations of the color fields sources in the nucleus,
$U(x_{\perp})$ is a matrix in the $SU(N)$ fundamental representation
which represents the interactions of the quark with the classical
color field of the nucleus.  All the information about the nature of
the medium crossed by the quark is included in the function
$C(l_T,x_2,A)$. In particular, it determines the dependence on the
saturation scale $Q_s$ (and on energy), implying that all saturation
effects are encoded in this function. In the Ref. \cite{MABMBGDCGC1}
we have shown that the saturation effects appear in the function
$C(l_T,x_2,A)$ only at small $l_T$, and as discussed here, the function
${\mathcal{W}} (p_T, l_T, x_1)$ selects values of $l_T$ larger than $p_T$,
implying that only dileptons with $p_T$ smaller than $Q_S$ should
carry information about the CGC.  Regarding the structure function
$F_2(x_1/z, M^2)$, the CTEQ6L parametrization \cite{Cteq6} was used and
the lepton pair mass gives the scale for the projectile quark
distribution.

The energy dependence introduced in the Eq. (\ref{eqcsh}) in the
correlator function $C(l_T,x_2,A)$ is performed by means of the
saturation scale $Q_s (x_2,A)$ and provides the investigation of the
effect of the quantum evolution in the dilepton production. We
employed the GBW parametrization \cite{GBW} to obtain the $x$
dependence of the saturation scale ($Q_s^2=(x)(x_0/x)^{\lambda}$), and
the parameters have been taken from the dipole cross section extracted
from the fit procedure by CGCfit \cite{IIMunier} parametrization.  The
nuclear radius, which appears in the Eq.  (\ref{eqcsh}), is taken from
the parametrization which has the form, $R_A=1.2 A^{1/3}$ $ fm$, while
the proton radius (for $pp$ calculations) is taken from the fit
\cite{IIMunier} ($R_p=0.6055 fm $).

The function $C(l_T,x,A)$ should be considered as a fundamental
quantity in the CGC formalism, since it carries all the information on
high density effects. We have evaluated the cross
sections using the function $C(l_T,x,A)$ from the mean-field
asymptotic solution for the JIMWLK evolution equation
\cite{Iancu:2001ad}, introducing an $x$ dependence through the nuclear
saturation scale, which is parametrized in the form $Q_s
^2(x,A)=A^{1/3}Q_s^2(x)$.  It results in a correlator function which
is a non-local Gaussian distribution of color sources.  These
considerations imply for the correlator the following form
\cite{Blaizot:2004wu,Gaussian}
\begin{eqnarray}
C(l_T,x,A) \equiv  \int d^2 x_{\perp} e^{il_T\cdot
  x_{\perp}}e^{\chi(x,x_{\perp},A)},
\end{eqnarray}
with
\begin{eqnarray}\nonumber
\chi(x,x_{\perp},A)&\equiv& -\frac{2}{\gamma c} \int
  \frac{dp}{p}(1-J_0(x_{\perp}p))\\
\times &&\ln\left(1+\left(\frac{Q_{2}^2(x,A)}{p^2}
\right)^{\gamma}\right),
\end{eqnarray}
where, $\gamma$ is the anomalous dimension ($\gamma\approx$ 0.64 for
BFKL) and $c\approx $ 4.84 \cite{Blaizot:2004wu,Gaussian}.

One has specified the cross section to evaluate the dilepton
differential cross section, and in the next section, the nuclear
modification ratio for the dilepton production is investigated and
related to the measured Cronin effect.

\section{Cronin effect and the dilepton production}

As discussed in the introduction of this work, the Cronin effect is
present in the measurement of the hadron transverse momentum
spectra. Here, the onset of the same effects in the dilepton
$p_T$ and rapidity spectra is investigated for a lepton pair mass
$M=6$ GeV. In a previous work \cite{MABMBGDCGC1}, the investigation
was performed for the $p_T$ distribution of dileptons with mass $M=3$
GeV.  Now, one evaluates the $p_T$ and rapidity spectra for the
dilepton at RHIC and LHC energies, $\sqrt{s}=200$ GeV and
$\sqrt{s}=8800$ GeV, respectively.

We have defined the nuclear modification ratio for the dilepton
production in the following form,
\begin{eqnarray}
R_{pA}=
\frac{\frac{d\sigma(pA)}{\pi R_A^2dMdydp_T^2}}{A^{1/3}
\frac{d\sigma(pp)}{\pi R_p^2dMdydp_T^2}}.
\end{eqnarray}
Some attention should be given to the uncertainty in the determination
of the nuclear radius, then each cross section is divided by the
nuclear or proton radius.  The factor $A^{1/3}$ was used in the
denominator to guarantee a ratio $R_{pA}$ about 1 at large $p_T$.

\begin{figure}[ht]
\scalebox{0.45}{\includegraphics{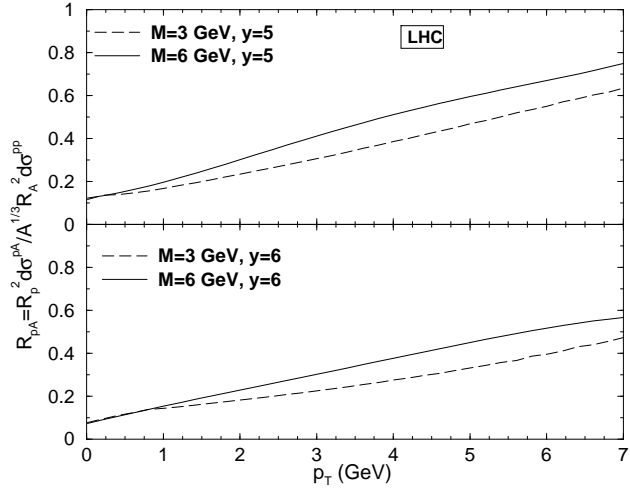}}
\caption{Ratio $R_{pA}$ for LHC energies, for $y=5$ and $y=6$,
comparing results for $M=3$ GeV and $M=6$ GeV.}
\label{ratioM6_3}
\end{figure}

The comparison between the results for the ratio $R_{pA}$ considering
two distinct lepton pair masses can be verified in the
Fig. \ref{ratioM6_3} for LHC energies, where the expected result is
found; the effect of the suppression of the ratio is reduced if the
dilepton mass is increased at a fixed rapidity. Such result can be
verified in the Ref. \cite{Kopeliovich}, where an analysis was
performed for the dilepton production at RHIC and LHC energies in the
color-dipole formalism, however, there the ratio was defined with a different
normalization.

This analysis regards forward rapidities, which maximum value depends
on the value of the mass and the transverse momentum. The region of
large mass and large $p_T$ implies smallest values for the rapidity
limit. At RHIC energies the maximum rapidity value reaches 4 and at LHC
energies it goes up to 7.

\begin{figure}[ht]
\begin{center}
\scalebox{0.85}{\includegraphics{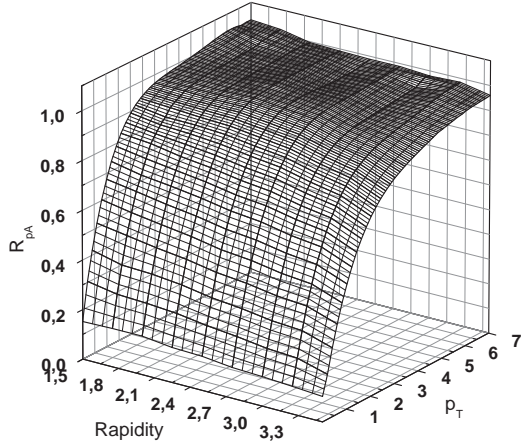}} 
\caption{Ratio $R_{pA}$ as a function of rapidity and $p_T$ for 
dileptons at RHIC energies.} \label{ratio3DRHIC}
\end{center}
\end{figure}

In the Fig. \ref{ratio3DRHIC} the nuclear modification ratio for RHIC
energies is shown for dilepton mass $M=6$ GeV.  A weak dependence of
the ratio $R_{pA}$ with the rapidity range is verified, since for a
fixed $p_T$ value, the ratio does not vary significantly with
rapidity. This occurs due to the fact that one evaluates the ratio
$R_{pA}$ only at forward rapidities, in a short limited range. For the
hadron spectra, the suppression of the ratio with the increase of the
rapidity is verified for a large range of rapidities, from $y=0$ to
forward ones ($y=3.2$) \cite{Arsene:2004ux}. In the case of the
dileptons, the same suppression should be verified, however the
calculation is restricted to the forward rapidities, providing a small
suppression in the rapidity range investigated here.  The suppression
of the ratio (absence of a Cronin type peak) in the $p_T$ distribution
is verified, although independent of the rapidity.

\begin{figure}[ht]
\scalebox{0.85}{\includegraphics{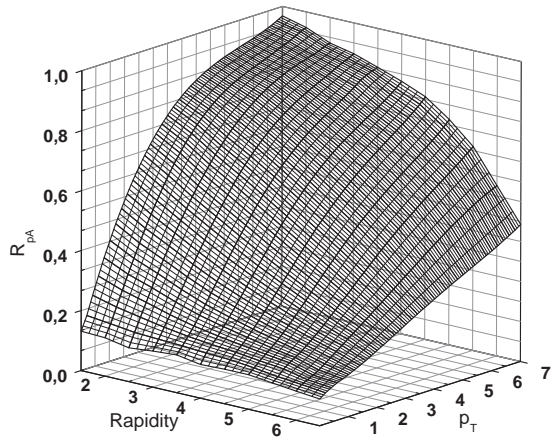}}
\caption{Ratio $R_{pA}$ as a function of rapidity and $p_T$ for dileptons
at LHC energies.}
\label{ratio3DLHC}
\end{figure}

In the Fig. \ref{ratio3DLHC} the nuclear modification
ratio for LHC energies is shown for dilepton mass $M=6$ GeV. Due to
the large range of forward rapidities at LHC energies, one verifies
the large suppression of the nuclear modification ratio with the
increase of the rapidity. This suppression is intensified at large
$p_T$.  The suppression of the same ratio with the transverse momentum
is also verified and is intensified at large rapidities.

At LHC, the large range of rapidity provides the $x$ range in the
large $p_T$ region ($p_T\approx 10$ GeV) between $10^{-4}$ and
$10^{-6}$. In this kinematical region, there are significant effects of
saturation predicted by the Color Glass Condensate: the large
suppression of the nuclear modification ratio comparing with the
expected Cronin peak shows the existence of the saturation effects, in
both, rapidity and transverse momentum distributions.

The predicted ratio for RHIC and LHC energies is evaluated considering
the same description for both, nucleus and nucleons. This implies that
the nucleon is not well described, since at intermediated $p_T$ the
nucleon is in the linear regime, and we are using a non-linear
approach to the proton. This simplification provides some uncertainty
in the ratio at large $p_T$, mainly at RHIC energies, where the proton
saturation scale reaches small values. At LHC energies, the
uncertainty in the ratio at large $p_T$ is reduced, since the
saturation scale of the proton reaches larger values.  The effect of
considering a realistic proton appears as a small reduction of the
ratio $R_{pA}$ at large $p_T$, since there is no large suppression
in proton gluon density in the linear regime. However, a realistic
description of the proton requires the determination of a new
parametrization, which is out of the scope of this work.

\begin{figure}[ht]
\begin{center}
\scalebox{0.85}{\includegraphics{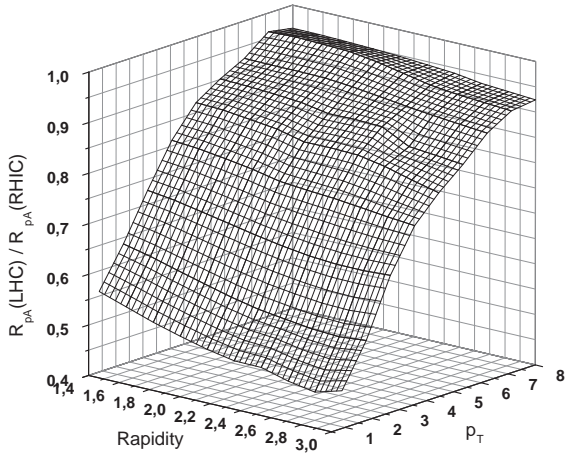}}
\caption{Ratio between LHC and RHIC ratio $R_{pA}$.}
\label{ratioLHCRHIC}
\end{center}
\end{figure}

The comparison between the LHC and RHIC results can be done by means
of the ratio $R_{pA} (LHC) / R_{pA} (RHIC)$. This ratio is presented
in the Fig. \ref{ratioLHCRHIC} which shows that the effect of the
suppression is enlarged at LHC, since it presents a severe reduction at
small $p_T$. We are restricted to the RHIC rapidity range to
provide this ratio, implying no strong suppression in the rapidity
spectra for the ratio $R_{pA} (LHC) / R_{pA} (RHIC)$ . The advantage
in analyzing such ratio is that it is less sensible to the proton
description at large $p_T$, emphasizing the CGC aspects we meant to
explore in this work.

In this work the ratio $R_{pA}$ was investigated in the framework of
the CGC for dilepton mass $M=6$ GeV and shown as a function of
rapidity and $p_T$, demonstrating the effect of suppression in both
distributions, rapidity and transverse momentum only at LHC energies.
One expects that such saturation effects are from a similar mechanism
as observed in the hadrons transverse momentum and rapidity spectra at
forward region. This contributes to clarify the status of the Cronin
effect as an initial state effect at forward rapidities. The ratio
between LHC and RHIC results shows significant saturation effects at
LHC compared with the RHIC ones. These features qualify dileptons
as a cleanest probe to the Color Glass Condensate dynamics at
forward rapidities. 

\section*{Acknowledgments} 

This work is supported by the CNPq, Brazil.

\end{document}